\documentclass[a4paper,11pt]{article}
\usepackage{pos}

\usepackage{caption}
\usepackage{subcaption}
\usepackage{hyperref}
\usepackage{url}
\usepackage{orcidlink}

\newcommand{\orcidauthorBENNETT}{0000-0002-1678-6701}
\newcommand{\orcidauthorLUCINI}{0000-0001-8974-8266}
\newcommand{\orcidauthorPIAI}{0000-0002-2251-0111} 
\newcommand{\orcidauthorFORZANO}{0000-0003-0985-8858}
\newcommand{\orcidauthorVADACCHINO}{0000-0002-5783-5602}
\newcommand{\orcidauthorHILL}{0000-0003-2383-940X}
\newcommand{\orcidauthorHONG}{0000-0002-3923-4184}
\newcommand{\orcidauthorDELDEBBIO}{0000-0003-4246-3305}
\newcommand{\orcidauthorLUPO}{0000-0001-9661-7811}
\newcommand{\orcidauthorLIN}{0000-0003-3743-0840}
\newcommand{\orcidauthorLEE}{0000-0002-4616-2422}
\newcommand{\orcidauthorZIERLER}{0000-0002-8670-4054}
\newcommand{\orcidauthorHSIAO}{0000-0002-8522-5190}

\title{Progress on the spectroscopy of lattice gauge theories using spectral densities}

\author[a]{Ed Bennett\,\orcidlink{\orcidauthorBENNETT}}
\author[b,c]{Luigi Del Debbio\,\orcidlink{\orcidauthorDELDEBBIO}}
\author*[d]{Niccol\`o Forzano\,\orcidlink{\orcidauthorFORZANO}}
\author[c]{Ryan C. Hill\,\orcidlink{\orcidauthorHILL}}
\author[e,f]{Deog Ki Hong\,\orcidlink{\orcidauthorHONG}}
\author[g]{Ho Hsiao\,\orcidlink{\orcidauthorHSIAO}}
\author[h]{Jong-Wan Lee\,\orcidlink{\orcidauthorLEE}}
\author[g,i]{C.-J. David Lin\,\orcidlink{\orcidauthorLIN}}
\author[a,j]{Biagio Lucini\,\orcidlink{\orcidauthorLUCINI}}
\author[k]{Alessandro Lupo\,\orcidlink{\orcidauthorLUPO}}
\author[d]{Maurizio Piai\,\orcidlink{\orcidauthorPIAI}}
\author[l]{Davide Vadacchino\,\orcidlink{\orcidauthorVADACCHINO}}
\author[d]{Fabian Zierler\,\orcidlink{\orcidauthorZIERLER}}

\affiliation[a]{Swansea Academy of Advanced Computing, Swansea University, Swansea, United Kingdom}
\affiliation[b]{Higgs Centre for Theoretical Physics, University of Edinburgh, Edinburgh, United Kingdom}
\affiliation[c]{School of Physics and Astronomy, University of Edinburgh, Edinburgh, United Kingdom}
\affiliation[d]{Department of Physics, Faculty of Science and Engineering, 
Swansea University, Singleton Park, Swansea, United Kingdom}
\affiliation[e]{Department of Physics, Pusan National University, Busan, Korea}
\affiliation[f]{Extreme Physics Institute, Pusan National University, Busan 46241, Korea}
\affiliation[g]{Institute of Physics, National Yang Ming Chiao Tung University, Hsinchu, Taiwan}
\affiliation[h]{Particle Theory and Cosmology Group, Center for Theoretical Physics of the Universe, Institute for Basic Science (IBS), Daejeon, Korea}
\affiliation[i]{Centre for High Energy Physics, Chung-Yuan Christian University, Chung-Li, Taiwan}
\affiliation[j]{Department of Mathematics, Faculty of Science and Engineering, 
Swansea University, Singleton Park, Swansea, United Kingdom}
\affiliation[k]{Aix-Marseille University, Universit\`e de Toulon, CNRS, CPT, iPhU, Marseille, France}
\affiliation[l]{Centre for Mathematical Sciences, University of Plymouth, Plymouth, United Kingdom}

\emailAdd{2227764@swansea.ac.uk}

\abstract{
Spectral densities encode non-perturbative information crucial in computing physical observables in strongly coupled field theories. Using lattice gauge theory data, we perform a systematic study to demonstrate the potential of recent technological advances in the reconstruction of spectral densities. We develop, maintain  and make publicly available dedicated analysis code that can be used for broad classes of lattice theories. As a test case, we analyse the $Sp(4)$ gauge theory coupled to an admixture of fermions transforming in the fundamental and two-index antisymmetric representations. We measure the masses of mesons in  energy-smeared spectral densities, after optimising the smearing parameters for available lattice ensembles. We present a summary of the mesons mass spectrum in all the twelve (flavored) channels available, including also several excited states. }

\FullConference{The 41st International Symposium on Lattice Field Theory (LATTICE2024)\\
 28 July - 3 August 2024\\
Liverpool, UK\\}

\begin{document}

\maketitle

\section{Lattice theory,  ensembles, and observables}

The analysis of spectral densities provides a novel tool to understand  non-perturbative aspects of  lattice gauge theories---see, e.g.,
Refs.~\cite{Hansen:2017mnd,Hansen:2019idp,
Bulava:2019kbi,
Bailas:2020qmv,
Gambino:2020crt,
Bruno:2020kyl,
Gambino:2022dvu,
DelDebbio:2022qgu,
Bulava:2021fre,
Kades:2019wtd,
Pawlowski:2022zhh,
DelDebbio:2021whr,
Bergamaschi:2023xzx,
Bonanno:2023thi,
Frezzotti:2023nun,Frezzotti:2024kqk}.
This proceedings contribution discusses our approach to reconstructing spectral densities using smeared correlation functions, focusing on the implementation of numerical techniques and their application to meson spectroscopy.
We exemplify the potential of such approach on the $Sp(4)$ gauge theory coupled to an admixture of fermions transforming in the fundamental and 2-index antisymmetric representations, which serves as a testbed for

exploring new physics scenarios, including composite Higgs models. We analyse new ensembles made available by
 the development of the research programme of \textit{Theoretical Explorations on the Lattice 
with Orthogonal and Symplectic groups} (TELOS)~\cite{Bennett:2017kga,Bennett:2019jzz,Bennett:2019cxd,Bennett:2020hqd,Bennett:2020qtj,Bennett:2022yfa,Bennett:2022gdz,Bennett:2022ftz,Bennett:2023wjw,Bennett:2023gbe,Bennett:2023mhh,Bennett:2023qwx,Bennett:2024cqv,Bennett:2024wda}---see also 
Refs.~\cite{
Kulkarni:2022bvh,
Bennett:2023rsl,
Dengler:2024maq,Bennett:2024bhy}.

The target theory of this study is the $Sp(4)$ gauge theory coupled to
 \(N_{({\rm f})} = 2\) fundamental and \(N_{({\rm as})} = 3\) antisymmetric fermions. We employ Wilson-Dirac fermions, with gauge configurations generated using an admixture of the  Hybrid Monte Carlo (HMC) and Rational HMC (RHMC) algorithms. The action on the lattice is written as
   $ S = S_g + S_f$,
where \(S_g\) is the Wilson plaquette gauge action, with coupling $\beta=8/g_0^2$, while \(S_f\) is the Wilson fermion action---for details, see Ref.~\cite{Bennett:2024cqv}. We assume the presence of two diagonal mass matrices for the two species of fermions, denoted as \(am_f^0\) and \(am_{as}^0\).
%
%
%
%
%
%
A summary of the parameters characterising the ensembles is provided in Table \ref{tab:ensembles}.  

We focus our attention on the twelve gauge invariant operators built as fermion bilinears, 
$\mathcal{O}(\vec{x}, t)$, with all the admissible spin structures, and off-diagonal flavor structure---for the singlets, see Ref.~\cite{Bennett:2024wda}.
One can extract the effective masses from correlation functions \(C(t)\), defined as
\begin{equation}
    C(t) = \sum_{\vec{x}} \langle 0 | \mathcal{O}(\vec{x}, t) \mathcal{O}^{\dagger}(\vec{0}, 0) | 0 \rangle\,.
\end{equation}
%
In order to improve the signal, 
we 
introduce APE~\cite{APE:1987ehd}  and Wuppertal~\cite{Gusken:1989qx} smearings,
and solve a Generalized Eigenvalue Problem (GEVP), 
to further
optimise the numerical quality of the ground state signal, as well as to detect excited states. 
We construct the operator basis by varying the smearing parameters.
Numerical results are listed in Tables~V to XVII of Ref~\cite{Bennett:2024cqv}.


\section{Spectral Density Reconstruction Method}

\begin{figure}[b]
    \centering
    \includegraphics[width=0.9\textwidth]{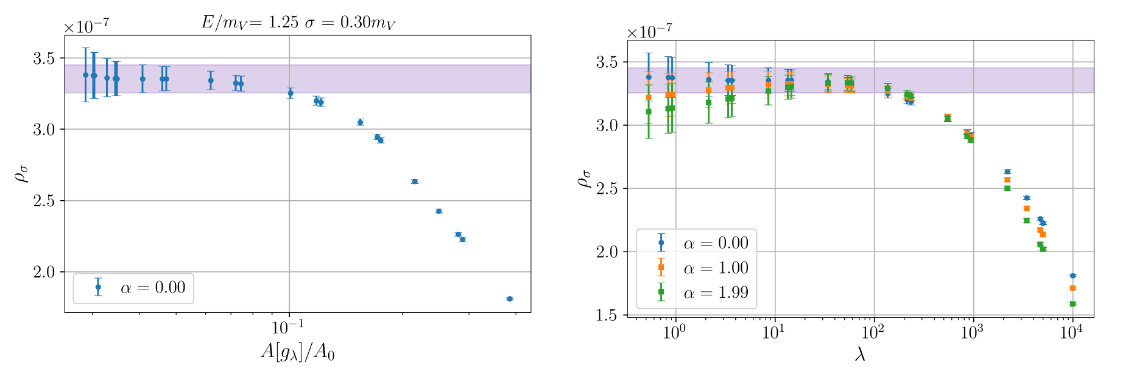}
    \caption{Examples of the optimisation of the spectral density reconstruction,
     for vector mesons (V)~\cite{Bennett:2024cqv}.
}
    \label{fig:plateau_finding}
\end{figure}

\begin{figure}[t]
    \centering
    \includegraphics[width=0.5\textwidth]{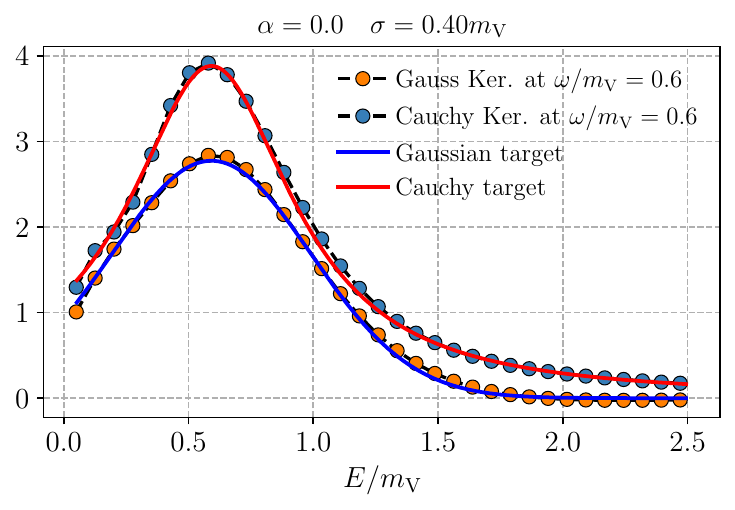}
    \caption{Spectral density kernels reconstructed with the HLT method, compared to target~\cite{Bennett:2024cqv}.}
    \label{fig:spectral_density_reconstruction}
\end{figure}

The spectral density, \(\rho(E)\), is  the inverse Laplace transform of the  correlation function, \(C(\tau)\):
\begin{equation}
    C(\tau) = \int_0^{\infty} dE \, \rho(E) e^{-E \tau}.
\end{equation}
We reconstruct it using the Hansen-Lupo-Tantalo (HLT) method~\cite{Hansen:2019idp}, a variation of the Backus-Gilbert method~\cite{Backus:1968svk}. 
To regularise this inversion, we introduce a smearing kernel \(\Delta_{\sigma}(E, \omega)\), which defines the smeared spectral density, \(\rho_\sigma(E)\):
\begin{equation}
\label{eq:mainformula}
    \rho_\sigma(\omega) = \int_0^{\infty} dE \, \Delta_{\sigma}(E, \omega) \rho(E).
\end{equation}
The parameter \(\sigma\) controls the smearing width, and therefore it controls the tradeoff between resolution and quality of the reconstruction. A larger \(\sigma\) broadens the kernel, reducing noise but blurring spectral features, while a smaller \(\sigma\) preserves fine details but increases statistical fluctuations.  
In order to balance the effects of statistical and systematic effects, we minimise, at fixed $\sigma$, a combined cost 
functional, $W[\vec{g}]$, defined as follows. We write the spectral density as $\rho_\sigma(E) = \sum_{\tau} g_\tau (E) \, C(\tau)$, the coefficients $\vec{g}=\{g_1,\,\cdots,\,g_{\tau_{\rm max}}\}$
corresponding to fixed-time lattice slices. We then define
\begin{equation}
    W[\vec{g}]\equiv A[\vec{g}] / A[0] + \lambda \, B [\vec{g}] / B_{\rm norm} ,
\end{equation}
where $A[\vec{g}] = \int_0^{\infty} d\omega \, e^{\alpha \omega} \left(\rho_\sigma(\omega) - \rho_{\text{target}}(\omega)\right)^2$, $B[\vec{g}] = \sum_{\tau \tau^{'}} g_\tau \, \mathrm{Cov}_{\tau \tau^{'}} [C] \, g_{\tau^{'}}$,  $B_{\rm norm} (E) = C^2(1)/E^2$,  \(\rho_{\text{target}}\) is the target spectral density 
extracted from the data,  
$\lambda$ is a trade-off parameter between systematic and statistical error and $\mathrm{Cov}[C]$ is the covariance matrix of the correlators $C(\tau)$. 
In principle, a particular choice of $\lambda$ introduces a source of bias that needs to be removed. Therefore, we perform a scan over $\lambda$ values, and we search for plateaus in the reconstructed spectral density. Figure~\ref{fig:plateau_finding} illustrates how the spectral reconstruction depends on $\alpha$ and $\lambda$. Having identified optimal values of these parameters, 
details of which can be found in Ref.~\cite{Bennett:2024cqv},
the minimisation of $W[\vec{g}]$ 
yields the coefficients, $\vec{g}$, and hence the reconstructed spectral density, $\rho_\sigma(E)$.
The process is repeated for each value of $E$ in the range of interest.

\begin{table}[t]
    \centering
    \caption{Summary table of the properties of the ensembles used in this study. The inverse coupling is denoted by \(\beta\), and the bare masses of the two species of fermions as \(am_f^0\) and \(am_{as}^0\), respectively~\cite{Bennett:2024cqv}.}
    \begin{tabular}{|c|c|c|c|c|}
        \hline
        Label & \(\beta\) & \(am_f^0\) & \(am_{as}^0\) & Lattice Volume \(\left(N_t \times N_s^3\right)\) \\
        \hline
        M1 & 6.5 & -1.01 & -0.71 & \(48 \times 20^3\) \\
        M2 & 6.5 & -1.01 & -0.71 & \(64 \times 20^3\) \\
        M3 & 6.5 & -1.01 & -0.71 & \(96 \times 20^3\) \\
        M4 & 6.5 & -1.01 & -0.70 & \(64 \times 20^3\) \\
        M5 & 6.5 & -1.01 & -0.72 & \(64 \times 32^3\) \\
        \hline
    \end{tabular}
    \label{tab:ensembles}
\end{table}

%



The systematic errors associated with the spectral density reconstruction are evaluated by varying the parameters \(\alpha\) and \(\lambda\). The first component of the systematic error is estimated as 
$\sigma_{1, \text{sys}}({\rho}(E)) = \left|{\rho}_{\lambda^*}(E) - {\rho}_{\lambda^*/10}(E)\right|$, 
and the second as $\sigma_{2, \text{sys}}({\rho}(E)) = \left|{\rho}_{\lambda^*, \alpha_2}(E) - {\rho}_{\lambda^*, \alpha_1}(E)\right|$,
where \(\lambda^*\) is defined by the optimisation procedure, and $\alpha_i$ are limiting values of 
$\alpha$. While most of the bias-removal comes from the scan over $\lambda$, by absorbing the bias into the statistical noise,  $\sigma_{1, \text{sys}}$ takes care of possible residual effect.


 Figure~\ref{fig:spectral_density_reconstruction} shows a comparison of the reconstructed smearing kernel, $\bar{\Delta}_\sigma (E, \omega) = \sum_{\tau} g_\tau (E) e^{-(t+1)E}$, with the target one,  for two choices of kernel. The  Gaussian kernel is
\begin{equation}
    \Delta_\sigma^{(\text{Gaussian})}(E, \omega) = e^{-\frac{(E - \omega)^2}{2\sigma^2}} / Z(\omega), \quad Z(\omega) = \int_0^{\infty} dE \, e^{-\frac{(E - \omega)^2}{2\sigma^2}}\,,
\end{equation}
while  the Cauchy kernel reads
\begin{equation}
    \Delta_\sigma^{(\text{Cauchy})}(E, \omega) = \frac{\sigma}{(E - \omega)^2 + \sigma^2}\,.
\end{equation}

\section{Meson Spectroscopy}

\begin{figure}[t]
    \centering
    \includegraphics[width=1.0\textwidth]{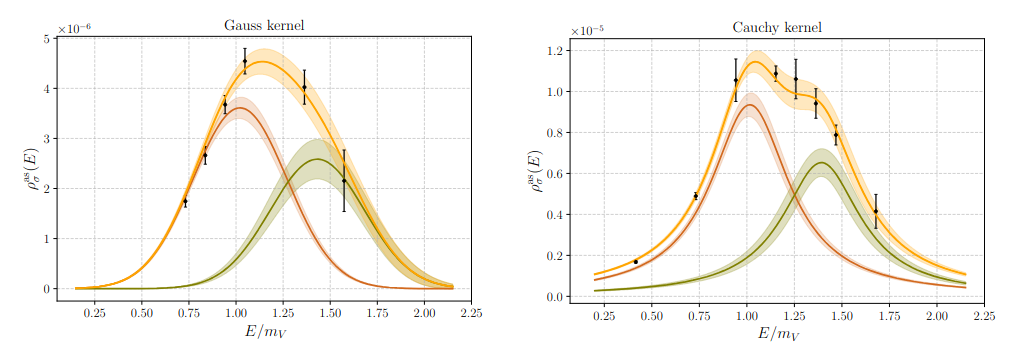}
    \caption{Reconstructed spectral density,  using Gaussian (left) and Cauchy (right) smearing kernels~\cite{Bennett:2024cqv}. 
    }
    \label{fig:smearing_kernels}
\end{figure}

The spectral densities, \({\rho}_\sigma(E)\), associated with  the meson correlation function, are fitted with both the Gaussian and Cauchy kernels, by
 minimising the functional:
\begin{equation}
    \chi^2 \equiv \sum_{E, E'} \left(f^{(k)}_\sigma(E) - {\rho}_\sigma(E)\right) \text{Cov}^{-1}_{E, E'}[{\rho}_\sigma] \left(f^{(k)}_\sigma(E') - {\rho}_\sigma(E')\right),
\end{equation}
where the fitting functions are, respectively,  $f^{(k)}_\sigma(E) = \sum_{n=1}^{k} A_n \Delta_\sigma^{(\text{Gauss})}(E - E_n)$, and $f^{(k)}_\sigma(E) = \sum_{n=1}^{k} B_n \Delta_\sigma^{(\text{Cauchy})}(E - E_n)$.
The  difference between energy levels determined with different kernels provides an estimate of 
systematic error, 
 $\sigma_{1, \text{sys}}(aE_n) = |aE_{n,\text{Gauss}} - aE_{n,\text{Cauchy}}|$. 
Figure~\ref{fig:smearing_kernels} shows a numerical example demonstrating the level of consistency: both ground and first excited states measured with the two kernels  are compatible, within statistical uncertainties.

\begin{figure}[t]
    \centering
    \includegraphics[width=1.0\textwidth]{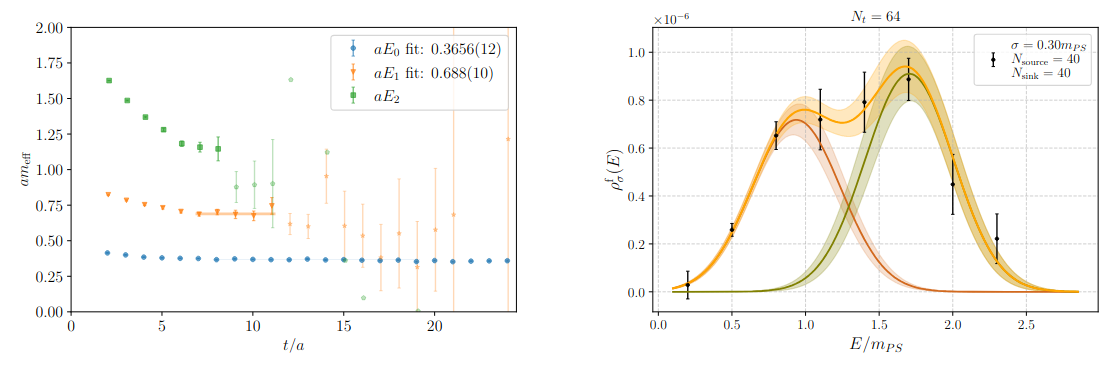}
    \caption{Comparison between GEVP (left) and  HLT (right) analysis used to measure ground state and excited state energy levels, for the pseudoscalar (PS) mesons~\cite{Bennett:2024cqv}.}
    \label{fig:energy_levels_comparison}
\end{figure}

\begin{figure}[t]
    \centering
    \includegraphics[width=0.7\textwidth]{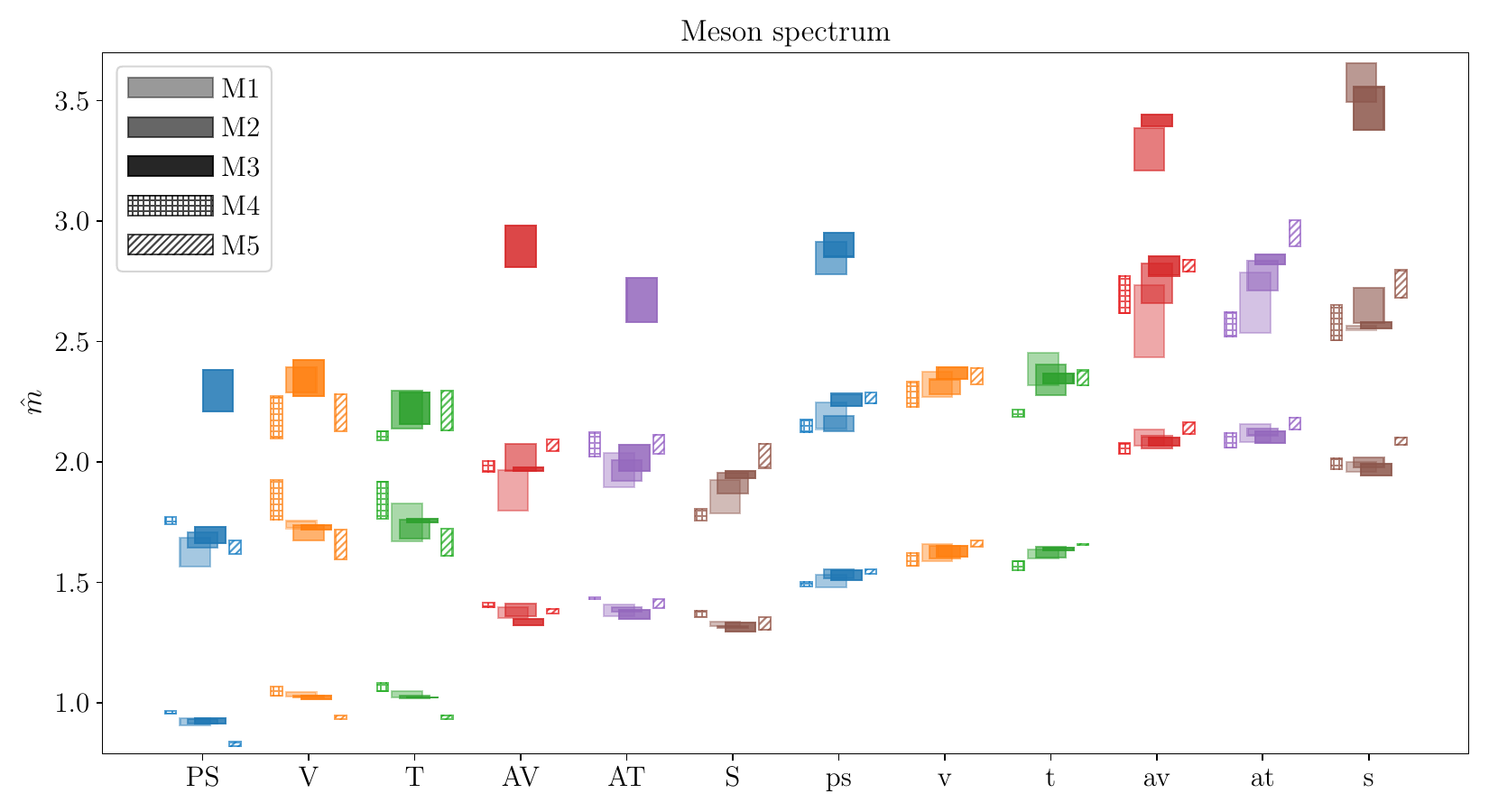}
    \caption{Mass spectrum of flavored mesons extracted with the  HLT spectral density reconstruction method~\cite{Bennett:2024cqv}.}
    \label{fig:meson_spectrum}
\end{figure}

The spectral density fitting method allows for a detailed exploration of excited states, which are often challenging. Figure~\ref{fig:energy_levels_comparison} shows a comparison between energy levels obtained with the GEVP and HLT methods, demonstrating consistency.
A comprehensive summary of numerical results for meson masses obtained with the HLT method are reported together with the GEVP results in Tables V to XVII of Ref.~\cite{Bennett:2024cqv}. 
The mesonic spectrum for the case study theory is displayed in Fig.~\ref{fig:meson_spectrum}.

\section{Outlook}
This study demonstrates the effectiveness of using smeared spectral densities,
by deploying the HLT method to the spectroscopy of flavored mesons in a special
 $Sp(4)$ gauge theory, used  as a case study. 
Future development will seek to apply these techniques to
other correlation functions, and to gain access to off-shell observables.
These technical developments
have the potential to impact of future studies of QCD as well as  new physics scenarios, offering new insights into the non-perturbative dynamics of strongly coupled theories.

\acknowledgments
This work has been supported  in part by 
the ExaTEPP projects EP/X017168/1 and EP/X01696/1,
the UKRI Science and Technology Facilities Council (STFC) Research Software Engineering Fellowship EP/V052489/1,
 the STFC Consolidated Grants No. ST/X508834/1, ST/P00055X/1, ST/T000813/1, 
  ST/X000648/1, and ST/P000630/1, a STFC new applicant scheme grant,
the Basic Science Research Program through the National Research Foundation of Korea (NRF) funded by the Ministry of Education (NRF-2017R1D1A1B06033701),
the National Research Foundation of Korea (NRF) grant funded by the Korea government (MSIT) (NRF-2018R1C1B3001379), 
the IBS under the project code, IBS-R018-D1,
 the National Research Foundation of Korea (NRF) grant funded by the Korea government (MSIT) (2021R1A4A5031460), the Taiwanese NSTC grant 112-2112-M-A49-021-MY3,
 the Royal Society Wolfson Research Merit Award WM170010,
  and  the Leverhulme Trust Research Fellowship No. RF-2020-4619,
 the European Research Council (ERC) under the European Union’s Horizon 2020 research and innovation program under Grant Agreement No. 813942. A.L is funded in part by l'Agence Nationale de la Recherche (ANR), under grant ANR-22-CE31-0011.
 
We used the Swansea SUNBIRD cluster (part of the Supercomputing Wales project) and AccelerateAI A100 GPU system. Supercomputing Wales and AccelerateAI are part funded by the European Regional Development Fund (ERDF) via Welsh Government. We also use the DiRAC Data Intensive service (CSD3) at the University of Cambridge, 
 the DiRAC Extreme Scaling service at The University of Edinburgh, and 
the DiRAC Data Intensive service at Leicester. The DiRAC Data Intensive service (CSD3) is managed by the University of Cambridge University Information Services on behalf of the STFC DiRAC HPC Facility. The DiRAC Data Intensive service at Leicester is operated by the University of Leicester IT Services, which forms part of the STFC DiRAC HPC Facility. The DiRAC Extreme Scaling service is operated by the Edinburgh Parallel Computing Centre on behalf of the STFC DiRAC HPC Facility (www.dirac.ac.uk). This DiRAC equipment is funded by BEIS, UKRI and STFC  capital funding and operation grants. DiRAC is part of the UKRI Digital Research Infrastructure. \\

{\bf Research Data Statement}---The data generated and analysis code used to prepare for these proceedings and the full paper \cite{Bennett:2024cqv} can be downloaded from  Refs.~\cite{data_release} and~\cite{analysis_release}. 

{\bf Open Access Statement}---For the purpose of open access, the authors have applied a Creative Commons  Attribution (CC BY) licence  to any Author Accepted Manuscript version arising.

\bibliographystyle{JHEP}
\bibliography{references.bib}

\end{document}